# Microwave magnetic dynamics in ferromagnetic metallic nanostructures lacking inversion symmetry

*M. Kostylev[1], Z. Yang[2], I.S. Maksymov[1,3], J. Ding[2], S. Samarin[1], and A.O. Adeyeye[2]*

[1]*School of Physics M013, The University of Western Australia, Crawley, WA 6009, Australia*
[2]*Department of Electrical and Computer Engineering, National University of Singapore, 117576 Singapore*
[3]*ARC Centre of Excellence for Nanoscale BioPhotonics, School of Applied Sciences, RMIT University, Melbourne, VIC 3001, Australia*

*Abstract:* In this work we carried out systematic experimental and theoretical investigations of the ferromagnetic resonance (FMR) response of quasi-two-dimensional magnetic nano-objects – microscopically long nanostripes made of ferromagnetic metals. We were interested in the impact of the symmetries of this geometry on the FMR response. Three possible scenarios, from which the inversion symmetry break originated were investigated: (1) from the shape of the stripe cross-section, (2) from the double-layer structure of the stripes with exchange coupling between the layers, and (3) from the single-side incidence of the microwave magnetic field on the plane of the nano-pattern. The latter scenario is characteristic of the stripline FMR configuration. It was found that the combined effect of the three symmetry breaks is much stronger than impacts of each of these symmetry breaks separately.

## I. Introduction

Ferromagnetic resonance (FMR) is a powerful tool for studying dynamic properties of metallic ferromagnetic thin films and nanostructures (see e.g., Refs. [1-12]). These materials are important for the emerging fields of magnonics [13], microwave spintronics [14-17], and for novel sensor applications [18-21].

Macroscopically-long stripes with sub-micron cross-section made of ferromagnetic materials have attracted a lot of attention, because this geometry is a very convenient model object for studying the impact of geometric confinement on magnetization dynamics on the sub-micrometre and nanometer scales [22-30]. The main reasons for the attractiveness of this geometry are the practical absence of a static demagnetizing field $\mathbf{H}_{ds}$ when the external field is applied along the stripes (i.e. along the axis *y* in Fig. 1(a)) and the quasi-two-dimensional (2D) character of the dynamics.

The former property allows clear separation of the static and dynamic demagnetizing effects. Furthermore, because of the absence of $\mathbf{H}_{ds}$, the static magnetization configuration for the stripes is very simple. The static magnetization vector has the same magnitude – equal to the saturation magnetization for the material - and the same direction – along the stripe – for every point on the stripe cross-section. Due to strong shape anisotropy for this geometry, this configuration remains very stable even at remanence [29-31]. The simplicity of the magnetization ground state, together with the 2D character of the magnetization dynamics, results in simple analytical and quasi-analytical theoretical models able to deliver a clear physical picture for the experimentally observed phenomena [23-24]. All this represents a clear advantage with respect to three-dimensional nano-objects, such as e.g. nano-discs, nano-squares of nano-rectangles, for which analytical models are very often inaccurate [32] and one needs to rely on brut-force numerical modelling using standard and (hence slow) software packages such as OOMMF [33], LLG [34] or muMAG [35] to get understanding, often limited, of the dynamics.

Experiment-wise, the break in the translational symmetry in the sample plane due to the nano-pattering makes it possible to probe standing spin waves with small wavelengths (< 300 nm or so) with such a simple tool as ferromagnetic resonance (FMR). This is due to the effective dipole pinning of magnetization at the stripe edges introduced by the nano-confinement [23]. This represents a big advantage with respect to continuous films, for which the static demagnetizing field is also absent but the translational symmetry is not broken. As a result of the higher symmetry, one needs to rely on such costly and sophisticated tool as Brillouin light scattering technique [36] to access those wavelengths for continuous films.

Historically, magnetic dynamics of single-layer magnetic stripes was studied first [22]. Theoretically [23], it was found that the stripes with rectangular cross-sections support a series of discrete standing spin wave resonances across the stripe width $w$, while the amplitude of precession is uniform across the stripe thickness $L$ for small ratios $L/w$. The formation of the discrete spectrum of the standing spin waves is possible because the typical stripe width ($w$<1μm) is significantly smaller than the spin wave free propagation path -at least several microns - for the typical magnetic material used in those experiments – $Ni_{80}Fe_{20}$ (Permalloy).

Later on it was realized that the magnetization dynamics on arrays of parallel stripes is collective [37,38], provided the edge-to-edge stripe separation is small enough. The collective character is due to coupling of stripes by magnetic dynamic dipole fields. This understanding led to suggestion of nano-scale magnonic crystals [39], as counterparts of the earlier suggested macroscopic ones [40].

Once the physics of the magnetisation precession on the single-layer stripes had been established, the interest shifted to multilayered stripes [24-27,30,41,42]. In parallel, drastic improvement of techniques of creating nanopatterned films made of ferromagnetic metals [43] made it possible to fabricate large arrays of nanostripes with sophisticated cross-sections [44].

Both abilities – to fabricate multi-layered nano-materials and to form sophisticated nano-shapes – made it possible to study the effect of symmetries on the magnetization dynamics. For the 2D materials, this issue has been addressed both theoretically and experimentally [30,44,45].

One more way to break symmetry of the system is to apply a symmetry breaking microwave magnetic field to an otherwise perfectly inversional-symmetric planar system [46]. This can be done in a simple broadband stripline ferromagnetic resonance (FMR) experiment [47], since in this case the microwave magnetic field is naturally incident only on one of the two surfaces of a planar nanostructure. The perfect shielding effect due to high conductivity of metals breaks the inversion symmetry. This leads to strong excitation of anti-symmetric standing spin wave modes across the confinement directions, which are the thickness [48] and the width [49] of nanoelements. The shielding is due to excitation of microwave eddy currents in the conductive materials. An important condition for existence of the eddy currents is continuity of the geometry in at least one in-plane direction. The geometry of very long nanostripes perfectly meets this criterion in the microwave frequency range.

The understanding of the impact of symmetries is far from being complete, and in this work we consider experimentally and theoretically a set of 2D geometries with increasingly lower inversion symmetries. The highest-symmetry system is an array of single-layered stripes of rectangular cross-section. The lowest symmetry system is an array of bi-layer nanostripes with an L-shaped cross-section and different thicknesses of the layers (Fig.1(a)). The latter were fabricated with an advanced technique recently suggested in [43].

We apply the microstrip FMR method to study the magnetization dynamics in these structures. Experimentally, we find that lowering the system symmetry increases the number of the observed FMR absorption peaks and leads to FMR traces of complicated shapes. Theoretically, we find that it is impossible to explain appearance of a significant number of high-amplitude peaks in the FMR spectrum, unless one assumes that there is an additional break of symmetry originating from the single-side incidence of the microwave magnetic field on the surface of the nano-pattern.

## II. Experiment

A number of bi-layered nanostripes have been fabricated by using the deep ultra-violet lithography. The lower layer was formed from $Ni_{80}Fe_{10}$ (Permalloy). It was capped by an iron (Fe) layer. The patterns were formed on silicon substrates. The width $w_1$ and the thickness $L_1$ of the Fe layer were varied from sample to sample. Also, reference single-layer stripes made of Permalloy (Py) and Fe were fabricated in the same process. The edge-to edge distance between the stripes on the array was 100nm.

The width $w_2$ and the thickness $L_2$ of the Py layer were always kept the same – 340nm and 10nm respectively. To allow $w_1<w_2$ (Fig. 1(a)), the recently developed tilted shadow deposition technique was employed [43]. It makes it possible to decrease the width of the formed nano-elements with respect to the width of the used photoresist lithographic pattern. Importantly, the formed nano-elements are located asymmetrically with respect to the photoresist pattern, since tilting the sample with respect to direction of the flow of the evaporated material breaks the symmetry of the electron-beam evaporation configuration.

Four samples with $w_1$=170 nm were fabricated with different Fe layer thicknesses $L_1$=2, 5, 10 and 20nm. Also, a sample with $L_1$=$L_2$=10nm and $w_1$=$w_2$=340nm ("rectangular stripe cross-section") was produced in the same production cycle. Its role was to serve as one more reference sample, allowing understanding of the role of the reduced width of the Fe layer on the magnetization dynamics of this complex system.

A representative SEM image of the fabricated bi-layer nanostripes with $w_1$=170nm, along with a sketch of the $w_1<w_2$ geometry ("L-shaped cross-section") is shown in Fig. 1. In this SEM image (Fig. 1(b)), one observes very well defined stripes with very well defined edges for the layers.

The left column of Fig. 2 displays exemplary frequency resolved microstripe FMR traces taken for some of these samples. The geometric details of these samples are listed in the second and the third columns of Table 1. The measurements were taken in a wide range of applied fields. It was found that there is no qualitative difference in FMR traces as long as the static magnetization configuration is the uniform "ferromagnetic" state – with the magnetization vector pointing along the stripes in the same direction for all stripes. To prepare this state the stripes were first magnetized to saturation in the longitudinal direction (i.e. along $y$) and then released from the saturation. To confirm the uniform magnetization ground state, FMR measurements were taken across the major hysteresis loops for the samples. The traces were typical for the ferromagnetic state of stripe arrays [31]. The latter traces are not important for the present study and therefore are not shown here. Given this, in the following we focus on FMR responses obtained for the ferromagnetic state at remanence (Figs. 2 and 3.)

Agilent PNA E8363C Microwave vector analyzer was employed to take the FMR measurements. To enable this, the samples, whose macroshape is a square with an area of 4x4 mm$^2$, were placed on top of a 30μm-wide coplanar line. The complex transmission coefficient (S21) for the microstrip line was measured as a function of the microwave frequency *f*. Fig. 2 (left column) and Fig. 3 display the absolute value of the measured S21 traces.

From Fig. 2 one sees that the number of absorption peaks increases with a decrease in the symmetry of the material. For the three single-layered materials (Rows A to C in Table 1, Figs. (a1)-(c1)) made of both Py and iron, the response is quasi-single-peak. We utilise the term "quasi" because in Fig. 2(a1) one more weak but narrow and well resolved peak is observed at about 7 GHz. A similar quasi-single-peak response is observed for the bi-layer stripes with rectangular cross-section (Row D of Table 1, Panel (d1) of Fig. 2). Strikingly, the peak frequency is larger than the one for the peaks in both Figs. 2(b1) and 2(c1).

For the bi-layer stipes with the L-shaped cross-section (Samples E and F from Table 1, Figs. 2(e1) and 2(f1)), the traces are very different from all the traces mentioned above: in addition to a strong peak located roughly at the same frequency as the large peak in Fig. 2(a1), one observes a dense family of peaks located slightly above the frequency of the peak from Fig. 2(c1). On closer inspection, one also finds that all the peak frequencies in Fig. 2(f1) are higher than for the respective peaks in Fig. 2(e1).

In Fig. 3 we demonstrate the evolution of the peak positions and amplitudes with an increase in the thickness of the Fe layer $L_1$ for L-shaped stripes. One sees that with the increase in $L_1$ the peak amplitudes and frequencies for the mode family gradually become larger. Also, the large low-frequency stand-alone peak gradually moves upwards in frequency.

Worth noticing are also peak widths (shown by the figures in the right-hand panels). To extract the widths, the experimental traces in Figs. 2 and 3 were fitted with the equation as follows:

$$S(f) = \mathrm{Re}\left( S^{(0)} + S^{(1)} f + \sum_{j=0}^{n} \frac{S_i}{f - f_j - i\Delta f_j} \right), \quad (1)$$

where *S*(*f*) is the fit to the experimental trace, *f* is the frequency, $S^{(0)}$ is the off-resonance level of the *complex* trace amplitude, $f_j$ is the resonance frequency for the *j*-th peak in the trace, $\Delta f_j$ and $S_j$ are its linewidth and *complex* amplitude respectively, *n* is the total number of peaks seen in the trace, *i* denotes the imaginary unit, and "Re" denotes the real part of the complex function in the brackets.

This formula follows from the result from [46]. In that paper it was found that the S21 value for a stripline loaded by a thin ferromagnetic film scales as $\exp(Z_r/Z_0)$, where $Z_0$ is the characteristic impedance of the stripline without the sample on its top and $Z_r$ is extra impedance inserted by the ferromagnetic resonance in the material. Equation (1) is obtained by taking a Taylor expansion of $\mathrm{Re}[\exp(Z_r/Z_0)]$ for $Z_r \ll Z_0$.

We also add an extra term - $S^{(1)}f$ - to this equation, in order to account for a linear slope seen in $S(f)$ in some traces off-resonance. ($S^{(1)}$ is also a complex number.)

This formula fits the experimental data very well. The fits reveal a tendency of an increase in the linewidth with the increase in $L_1$. Interestingly, all the resonance lines for the L-shaped stripes are significantly narrower than observed for the single-layer reference iron samples (Fig. 2(b1) and 2(c1)) – 831 and 819 MHz, respectively. However, most of them are broader than the main peak for the bi-layer stripes with the rectangular cross-section (Fig. 2(d1)) – 172 MHz.

## III. Theory

### A. The numerical model

The goal of our theory is to explain the appearance of the family of modes seen in the experimental traces for the L-shaped stripes and which is absent for the bi-layer stripes with the rectangular cross-section Fig. 2(d1). We also strive to explain the observed behaviour of the resonance linewidths.

To this end we carry out calculations of stripline FMR absorption by the samples. We employ an original numerical code which is a further development of the quasi-analytical approach from [44] and [49]. It is based on solution of the linearized Landau-Lifshitz-Gilbert (LLG) equation in the magnetostatic approximation.

Magnetization dynamics of the very long stripes driven by the quasi-uniform field of the microstrip FMR setup is two-dimensional (2D). This makes the theoretical description simple and the numerical code very fast. We employ a 2D Green's function description of the dynamic dipole field of the precession magnetization [44]. Since the stripe array is dense (stripe edge-to-edge separation is 100nm) the dynamic dipole field of the array is collective [37]. The collective behaviour is also included in the theory, although it is not important for understanding of the physics we aim to explore in this work.

Since the sizes of the stripe cross-section are comparable with the exchange length for Permalloy (5nm) we also include contribution from the effective exchange field of precessing magnetization. We assume "unpinned surface spins" boundary conditions at all stripe surfaces [50].

Both the integral operator involving the Green's function and the differential operator of the effective exchange field are discretized on a square mesh which fills the stripe cross-section. This procedure is the same as previously employed in [44, 49]. The important peculiarity of the present geometry is the presence of two different ferromagnetic materials, potentially in exchange contact through the interface between them. Therefore the model assumes that the values for the saturation magnetization, intra-layer exchange constant, gyromagnetic coefficient, and Gilbert damping constant are different above and below the layer interface in Fig. 1(a). On top of this, the interface exchange boundary conditions for the dynamic magnetization are included in the model. We use the linearized boundary conditions from [51]:

$$\begin{aligned} &\partial m_x^{(1)}/\partial z + \frac{A_{12}}{A_1} m_x^{(1)} - \frac{A_{12}}{A_1}\frac{M_1}{M_2} m_x^{(2)} = 0 \\ &\partial m_x^{(2)}/\partial z - \frac{A_{12}}{A_2} m_x^{(2)} + \frac{A_{12}}{A_2}\frac{M_2}{M_1} m_x^{(1)} = 0 \end{aligned} . \qquad (2)$$

In this expression, $m_x^{(1)}$ and $m_x^{(2)}$ are the in-plane ($x$) components of the dynamic magnetization in Layers 1 and 2 respectively, $A_1$ and $A_2$ are intra-layer exchange constants for the layers, $A_{12}$ is the inter-layer exchange constant, and $M_1$ and $M_2$ are saturation magnetization values for the layers. A similar boundary condition is used for the perpendicular-to-plane (i.e. perpendicular-to-the-interface) component of dynamic magnetization $m_z$. As in [46], the surface and interface boundary conditions are included into the boundary elements of the discrete (finite-difference) version of the exchange operator.

At this stage, the microwave driving field is not included into the model. Thus, we are solving numerically a homogeneous integro-differential equation which follows from the linearized LLG equation. The discretization of the operators transforms the equation into a generalized eigenvalue problem for a pair of matrices:

$$i\omega\hat{\Lambda}\,|\,m>=\hat{C}\,|\,m>. \quad (3)$$

Here $\omega=2\pi f$, $|m>$ is a block column-vector with blocks $(m_{xj}, m_{zj})$, where $m_{\lambda j}$ is the λ-component (i.e. $x$ or $z$ component) of dynamic magnetization for the point $j$ of the mesh, $\hat{\Lambda}$ is a matrix which follows from the Gilbert term of the LLG equation, $\alpha_G$ is the Gilbert damping constant, and $\hat{C}$ is the matrix which is obtained from the precessional part of the LLG equation. Since $\hat{\Lambda}$ is a diagonal block matrix, its inverse $\hat{\Lambda}^{-1}$ is easily obtained analytically. This results in an eigenvalue/eigen-vector problem

$$i\omega\,|\,m>=\hat{C}'\,|\,m>. \quad (4)$$

where $\hat{C}'=\hat{\Lambda}^{-1}\hat{C}$. The diagonal blocks of $\hat{\Lambda}^{-1}$ are given by:

$[\hat{\Lambda}^{-1}]_{j,j}=\begin{bmatrix} 1 & -\alpha_G^{(\lambda)} \\ \alpha_G^{(\lambda)} & 1 \end{bmatrix}$, where $\alpha_G^{(\lambda)}$ is the value of the Gilbert constant for Layer $\lambda$ ($\lambda$=1 or 2, depending on the mesh point $j$). To derive this expression we took into account that $\alpha_G^{(\lambda)}<<1$.

To obtain numerical solutions for the eigen-value problem for the matrix $\hat{C}'$ we use the numerical tools built in MathCAD. For the mesh size $3.33\times3.33\text{nm}^2$, it takes about 5 minutes to obtain all eigenvalues and eigenvectors of $\hat{C}'$ with a quad-core personal computer for the most computation-demanding geometry – that of the L-shaped stripes with the 20nm-thick Fe layer.

Once the solutions of the homogeneous problem have been obtained, the inhomogeneous vector-matrix equation

$$i\omega\,|\,m>-\hat{C}'\,|\,m>=\hat{\Lambda}^{-1}U\,|\,h_{dr}> \quad (5)$$

is solved, by expanding the solution using the basis of the obtained eigen-vectors. Following [49], an effective microwave magnetic susceptibility $\kappa$ may be introduced as

$$\kappa(\omega) = \frac{1}{< h_{dr} \mid h_{dr} >} \sum_j \frac{< h_{dr} \mid m_j^0 >< m_j^0 \mid \hat{\Lambda}^{-1} U \mid h_{dr} >}{\omega - \omega_j}, \quad (6)$$

where $| m_j^0 >$ is the eigenvector of $\hat{C}'$ corresponding to its *j*-th eigenvalue $\omega_j$, $< m_j^0 |$ is the corresponding eigen-vector of a matrix transposed to $\hat{C}'$, $| h_{dr} >$ is a vector containing amplitudes of the driving field at the mesh points, $< h_{dr} |$ is a vector transposed to it, $\omega$ is the frequency of the driving field, $< \cdots | \cdots >$ denotes scalar product of two vectors, and $U$ is a diagonal block matrix with blocks $U_{j,j} = \gamma_\lambda M_\lambda \begin{bmatrix} 0 & 1 \\ -1 & 0 \end{bmatrix}$ and $\gamma_\lambda$ is the gyromagnetic coefficient for Layer $\lambda$. Note that $< m_j^0 | m_{j'}^0 >= \delta_{j,j'}$, where $\delta_{j,j'}$ is Kronecker Delta, and that the resonance eigen-frequencies $\omega_j$ are complex values, because magnetic damping is included into the model. Also note, that there is no need to solve the additional eigenvector problem for the transposed matrix in order to calculate the set of $< m_j^0 |$. These vectors are obtained from the solution for $| m_j^0 >$ with a simple procedure of inverting a matrix containing all $| m_j^0 >$ vectors as its columns [49].

## B. Most general details of the theory of the magnetization dynamics in layered nanostripes

The theoretical understanding of the magnetization dynamics in single-layer [23,37] and bi-layer [24,25,27,41,42,45] stripes with rectangular cross-section has been well established. As already mentioned in Introduction, the single-layer stripes are characterized by a number of standing spin wave resonances across the stripe width [23].

Dynamics of bi-layer stripes is more complicated. Strongly exchange-coupled ($a\ A_{12} \cong A_1, A_2$, where $a$ is the size of the crystal unit cell) and completely exchange-uncoupled bilayer stripes ($A_{12}$=0) behave differently. For any strength of the exchange coupling through the interface, the standing spin wave resonances separate in two families of modes: the acoustic and the optical resonances [24,25]. For the acoustic resonances the magnetization vector precesses in-phase for both layers and for the optic mode there is a difference of π between the phases of the precession. Importantly, for the bi-layer stripes the dynamic dipole field configuration is such that the anti-phase layer coupling reduces the total energy (i.e. the energy contribution is negative), but the in-phase one results in a positive contribution to the total energy. As a result, the dipole energy for the acoustic mode is higher than for the optical one.

Resonances of a bi-layer system are formed on the base of the resonance frequencies for single-layer stripes with the same cross-section sizes as for the layers of the bi-layer stripe (“parent stripes”). In the most symmetric case of two layers having the same thickness and made of the same material, the parent stripes are the same for both layers. Therefore, in the absence of exchange coupling of the layers, a pair of acoustic and optical resonances is formed on the base of the same resonance frequency. Given the sign of the coupling field contribution to the total energy (see above), the optical mode of the pair is characterized by a lower resonance frequency with respect to the acoustic one.

Alternatively, one may think of the acoustic modes as modes of an effective single-layer stripe with the thickness equal to the total thickness of the bi-layer system [24,26]. An increase in the thickness of a single-layer stripe increases its resonance frequencies [23]. Hence the acoustic mode should have a larger frequency and, consequently, the optical one a smaller frequency than the respective parent frequency. Usually the difference in the two frequencies is large for all pairs. This results in separation of modes in the two distinct mode families.

However, the separation is not complete. For both mode families the resonance frequency increases with the mode number, i.e. with the decrease in the wavelength of the standing spin wave. This implies that some overlap in frequency ranges for the two mode families is possible – the frequencies of some higher-order frequency modes of the optical family may be higher than the lowest frequency for the acoustic family (see, e.g., Fig. 5 in [25]).

Different thicknesses for the two layers or different magnetic materials for them break the symmetry. Physically, the dynamic dipole-dipole interaction should result in mode repulsion (i.e. mode conversion is physically prohibited). Therefore, the family of (quasi)-optic modes should now be formed on the base of the resonances in the parent stripe with smaller eigen-frequencies and the quasi-acoustic resonances should be formed on the base of the resonances in the parent stripe with higher eigen-frequencies. For instance, an increase in the parent stripe thickness shifts the resonance frequencies upwards. Hence, the acoustic mode family (as located higher in frequency) will be formed on the base of the resonances in the thicker parent stripe and the optical family (as located lower in frequency) on the basis of the resonances in the thinner stripes [30]. Similarly, for two different saturation magnetization values but the same thicknesses of the layers, the acoustic resonances will originate from the family of modes of the parent stripe with the larger saturation magnetization [25,45].

Importantly, the strength of coupling decreases with an increase in the frequency difference for the resonances in the parent stripes. The larger the difference in thicknesses or saturation magnetizations, the smaller the coupling. One important consequence of this is that the dynamics of a coupled system with broken symmetry tends to more and more concentrate in one of the layers with an increase in the system asymmetry. For instance, the precession amplitude for the acoustic mode becomes larger and larger for the thicker layer and smaller and smaller for the thinner one with an increase in the difference $L_2$–$L_1$. Equivalently, the amplitudes of the optical resonances are larger in the layer with a smaller $M$ value and the amplitudes of the acoustic modes are larger in the layer with larger saturation magnetization. This is well seen in Fig. 6 from Ref. [25].

The ferromagnetic resonance is the microwave response of a material to application of a spatially uniform microwave magnetic field. The uniformity of the driving field implies that the response amplitude for a particular resonance mode scales as the net dynamic magnetic moment for the mode, .i.e. as the mode amplitude averaged over the material volume. This implies that the FMR responses of all optical modes for the bi-layer stripes consisting of two identical layers vanish. Only responses of in-plane symmetric acoustic modes will be seen in the FMR trace, with the peak amplitude dropping quickly with the mode number. The same applies to the single-layer stripes: only standing-wave modes which are symmetric in the direction of the stripe width (along $x$) contribute to the FMR absorption. Again, the peak amplitude decreases quickly with an increase in the peak number. This is clearly seen from Fig. 2(a2) –2(c2), where we employed the theory from [49] to produce the traces.

The break in the system symmetry makes the optical resonances in bi-layer stripes observable due to unequal precession amplitudes in the layers. The larger the difference in the layer parameters, the larger amplitude of the FMR response of the optical resonances may be expected.

Interface exchange coupling between the layers modifies the spectrum of the resonance modes. It shifts the frequencies of the optical mode family upwards. The larger the interface exchange constant, the larger the shift [27]. As a result, the optical modes may now exist in the same or even higher frequency range than the acoustic modes. This frequency shift is due to the extra intra-layer exchange contribution to the resonance frequencies for the optical resonances.

The shift may be understood by considering the same illustrative case of two identical layers. In the limiting case of the inter-layer exchange constant equal to the intra-layer exchange constant ($aA_{12}=A_1$), the optical resonances of the bi-layer should reduce to the fist (anti-symmetric) standing spin wave ($1^{st}$ SSW) across the *thickness* of an equivalent single-layer stripe of double thickness. Similarly, the acoustic resonances of the bi-layer reduce to the main (thickness-uniform) family of resonances of the equivalent single-layer stripe. Since $L_1+L_2<<w$, the resonances formed on the base of the $1^{st}$ SSW should have much higher frequencies than the ones of the main mode family.

Similarly, in the case of two exchange-coupled layers with different $M$ values, one may think of the acoustic resonances as resonances in an effective single-layer stripe with a thickness $L_1+L_2$ (or close to it) having a value of $M$ lying somewhere between $M_1$ and $M_2$. The optical resonances then may be thought as the ones based on $1^{st}$ SSW of the same equivalent single-layer stripe whose frequency positions will be determined by an effective exchange constant. Obviously, this model should become less and less valid with an increase in the difference in the layer parameters and/or with a decrease in the strength of the interface exchange interaction. This is because those changes should result in larger localization of oscillations for particular resonance modes in particular layers, similar to the above discussed case of the purely dipole-dipole coupling of the layers.

Further exploiting the same similarity, one may arrive at a conclusion that the increase in the strength of the interface exchange coupling should decrease the amplitude of the FMR response of the optical modes and increase the amplitude of the acoustic ones. However, this conclusion is valid only for magneto-insulating materials or for FMR measurements of conductive materials taken with a microwave cavity. Our samples are made of highly conducting metals. Therefore, in the conditions of the stripline FMR one may expect strong non-uniformity of the microwave magnetic field over the stripe cross-section [46,49] due to the microwave shielding effect. This effect arises because the single-side incidence of the microwave magnetic field on the sample surface breaks symmetry of the microwave field geometry. This results in excitation of microwave eddy currents in the materials (see Fig. 21 in Ref. [47] for the explanation).

In our case of bi-layer materials the shielding effect is accentuated by a significant difference in layer conductivities. The tabular value of Fe conductivity is about two times larger than that of Permalloy. Furthermore, the Fe layer overlays the Permalloy one. Then, from the analogy with Fig. 1(b) in [52] one may expect that the eddy currents and hence the total microwave magnetic field is larger in the Fe layer than in the Permalloy one. Since the magnetization precession in metallic materials is driven by the total microwave magnetic field [46], one may expect that the microwave driving field is stronger in the Fe layer than in the Permalloy one.

In Ref. [49] we suggested a method how to estimate the cross-sectional distribution of the total microwave magnetic field for single-layer metallic stripes. In this work we extend this theory to the case of bi-layer stripes of rectangular cross-section (see Appendix). We will use this result below to simulate the FMR response for the geometry of Fig. 2 (d1).

However, this approach loses its accuracy for the L-shaped stripes. Therefore, in order to get understanding of the eddy-current contribution to the microwave field configuration for the L-shaped stripes, we carried out a separate numerical simulation employing original Finite-Difference-Time-Domain (FDTD) software [53]. The FDTD is a numerical analysis method used for solving complex problems of computational electrodynamics by finding approximate solutions to the system of the differential Maxwell's equations. This method is chosen because it allows one to calculate the electromagnetic field distribution in systems consisting of materials having disparate electrical conductivities, which is the case of the investigated L-shaped stripes. It is also noteworthy that a rectangular finite-difference mesh of the FDTD method easily aligns with the surfaces of the layers of the L-shaped stripes. Together with the periodicity of the investigated L-shaped structures, which dramatically simplifies simulations by imposing periodic boundary conditions, it makes it possible to obtain high accuracy numerical results while keeping computational efforts within reasonable limits.

Figure 4 displays the calculated profile of the total microwave magnetic field averaged over the thickness of the Fe and Py layers. In this simulation it is assumed that the spatially uniform microwave magnetic field of the microstrip line is incident from the side of the Fe layer. From this figure one sees that the in-plane component of the total field has noticeably larger amplitude in the Fe layer. Furthermore, one also notices significant reduction in the field amplitude in the section of the Permalloy layer not capped by Fe. Hence, a strong eddy current in Fe layer shields not only the capped section of the Permalloy layer, but also induces microwave shielding effect in a large area around the Fe stripe.

We have gone this long way of discussing the details of the microwave magnetic dynamics and electrodynamics of nanomaterials of interest, because, given the large number of parameters involved, it is impossible to get agreement with the experimental data, without employing some a-priori knowledge. This point will become clear from the next section.

### C. Computational results: FMR peak frequencies and amplitudes

The results of our computations are displayed in the right-hand panel of Fig. 2. One sees reasonable agreement with the experiment. Table 1 shows the values of magnetic parameters for the materials which provide the best agreement of the simulation results with the experimental data.

Instrumental for understanding of the experimental graphs from Figs 2 and 3 is Fig. 2 (d1). The discussion above suggests that, if the layers in the simple geometry of the rectangular stripe cross-section were not exchange-coupled or coupled only weakly, one should observe a second large-amplitude peak– the fundamental optical mode, see the dashed and the dotted lines in Fig. 2(d2). In the case of the completely exchange-uncoupled layers the fundamental optical mode would be easily identifiable – it would be located below the position of the large peak in Fig. 1(a1), see the dashed line in Fig.2 (d2).

The absence of such a prominent peak in the experimental data suggests that the layers are strongly exchange coupled. The coupling shifts the fundamental optical peak (the smaller peak in the trace shown by the dotted line) beyond the upper limit of the experimentally probed frequency range. The result of a simulation carried out for a large value of the interface exchange constant (solid line in the same figure) is in good agreement with the experiment.

The simulation identifies the large peak in the trace as the fundamental acoustic mode. Its dynamics is very similar to one of an effective 20nm-thick single layer stripe. The larger thickness shifts the peak upwards with respect to the peak in Fig. 2(a2). So does the larger saturation magnetization of the effective stripe with respect to the one for Permalloy. The combined effect of both frequency shifts results in the peak frequency close to one for the peak from Fig. 2(c1). Here one has to note that the aspect ratio $(L_1+L_2)/w_1$ for the geometry from Fig. 2(d2) is the same as for Sample C (Fig. 2(c2)). As the dynamics of the lowest-order mode for single-layer stripes is mostly determined by the dipole-dipole interaction, whose strength scales as the aspect ratio, it is not surprising that the peak frequency from Fig. 2(d2) is close to one from Fig. 2(c2).

Based on the analysis from the two preceding paragraphs, it is natural to assume that the Fe and Py layers are strongly exchange coupled for all bi-layered samples we fabricated. Therefore, the simulations for the L-shaped stripes were carried out assuming the same value of the inter-layer exchange constant as for Fig. 2(d2). The simulation results are displayed in Panels (e2) and (f2) of Fig. 2. One sees good qualitative agreement with the experiment - the number of prominent peaks and there relative amplitudes are the same in the experimental and the theoretical graphs.

Here one has to mention that in our simulations we were unable to use the same saturation magnetization, intra-layer exchange constant, and Gilbert damping constant values for all right-hand panels of Fig. 2 (see Table 1). Most importantly, it was found that it is impossible to simulate the data in Fig. 2(d2) employing the respective parameters from the first and the third rows of the table (Samples A and C). In the first place, our computations showed that such a narrow resonance line as in Fig. 2(d2) cannot be obtained for the value of the magnetic loss parameter for Fe used to produce the traces in Fig.2(b2) and 2(c2). One needs a much smaller value to obtain the resonance line as narrow as in Fig.2(d2). Hence the theory suggests that a Fe layer grown on top of the Permalloy one may be of better quality than grown on a bare Si substrate (Fig.2(c2)).

This conclusion of potential difference in magnetic parameters for the layers of the layered structures from the ones for the reference single-layer stripes justifies our usage of individual values of material parameters for each individual sample. One sees that there is some trend in the values from Table 1, for instance, one of larger $M_1$ and $M_2$ for the bi-layer structures than for the reference single-layer stripes. However, no well defined correlations are observed. Therefore, we claim that our theoretical model may not include one or several important magnetic parameters for the materials. One of them may be interface and surface anisotropies. Since the anisotropies are neglected, they may renormalize the values of parameters which are included in the model. The “renormalization” means that the parameter values from Table 1 which result in the best fit of the experimental data are only effective values. Given that the resonance frequencies scale differently as functions of different parameters, the effective values are valid for a specific sample only, and, potentially, only for a specific value of the applied static field.

The anisotropies are not included in the computations just in order to minimize the number of fitting parameters and hence the number of program runs required to explore the space of model parameters. For instance, inclusion of anisotropy for just one layer would require adding at least 3 extra parameters – type and strength of the anisotropy, and the anisotropy axes directions.

The good qualitative agreement between the theory and the experiment for the L-shaped stripes (Panels (e) and (f) of Fig. 2) allows us to explain the formation of the family of peaks denoted as "2","3" and "4" in the theoretical traces. To this end we carry out calculation of the profiles of the modes responsible for those peaks. Figure 5 displays the dynamic magnetization distributions (mode profiles) for modes designated by respective figures in Figs 2(e2) and 2(f2). The solid lines show the amplitude of the in-plane ($x$) component of dynamic magnetization averaged over the thickness of the Fe layer. The dashed ones are the respective profiles for the Py layer. One sees that all these modes belong to the same family of acoustic modes. This is evident from the same sign for the dynamic magnetization amplitude in both layers for the part of the stripe cross-section where the Permalloy layer is buried under the iron layer ($0< x<170$nm, i.e. to the left of the vertical dashed line). One also sees that for both Fe layer thicknesses the modes contributing to the FMR response are the same.

The largest peak in Figs. 2(e2) and (f2) corresponds to the fundamental acoustic mode. Its profile is shown in Panels (e2)-1 and (f2)-1 of Fig. 5). On closer inspection of these mode profiles one notices that the amplitude of magnetization precession in the Permalloy layer is nearly negligible for $0< x<170$nm. Furthermore, its shape is similar to an exponential function rather than to a symmetric harmonic function which one might expect for a standing-wave resonance. The exponential profile implies that the precession in Permalloy for $0<x<170$nm is more likely forced, rather than represents an eigen-oscillation. So does the precession in the Fe layer which is evidenced by a significant asymmetry of the solid lines in these panels. The precession for $0<x<170$nm is mainly driven through exchange interaction via a "virtual interface" $x=170$nm by the standing wave resonance taking place in the part of the Permalloy layer not covered by Fe ($170\text{nm}<x<340$nm).

With an increase in the peak number, the amplitude profiles for both layers for $0<x<170$nm become more and more symmetric. Also, the magnetization precession amplitudes for $0<x<170$nm become larger with respect to the amplitudes in the part of the Permalloy layer not covered with Fe ($170\text{nm}<x<340$nm). The increase may be understood as a gradual change in the character of precession in the left-hand part of the stripe cross-section ($0<x<170$nm). The precession becomes less and less forced and more and more resonant with the increase in the mode number. This happens because the peak frequency increases with the increase in the mode number. The frequency thus becomes closer to the one for the fundamental mode of a single-layer 170nm-wide Fe stripe with $4\pi M_1=18500$ G (see the respective values in the two last rows of Table 1.) This changes the precession character.

The same change in the oscillation character is responsible for the gradual increase in the height of the resonance absorption peaks with the increase in the peak number. Here one has to note that the amplitudes of Peaks 2, 3 and 4 with respect to the amplitude of Peak 1 ("relative amplitude") strongly depend on the assumed profile for the microwave driving field employed for calculations with Eq.(6). If one assumes that the microwave magnetic field is spatially uniform, one obtains the trace shown by the dashed line in Fig. 2(f2). One sees that the relative amplitudes of the mode family are significantly smaller in this case than the experimental ones. However, if one uses the profile from Fig. 4 as the assumed distribution of the microwave driving field, one obtains the correct ratio of the amplitudes of the higher-order modes to the amplitude of the fundamental acoustic peak in the calculation.

Hence, the microwave shielding effect plays a very important role in the formation of the stripline FMR response of these materials. Recall, that this effect arises because of the single-side incidence of the microwave field on the sample in the conditions of the stripline FMR. Thus, the system is strongly affected by one more break of symmetry – that of the inversion symmetry of the excitation geometry. The break of the latter does not significantly affect the bi-layer stripes with the rectangular cross-section (Fig. 2(d2)), because of higher symmetry of this sample geometry. However, when the inversion symmetry of the sample cross-section is further reduced by the introduction of the $L$-shape, the combined effect of three inversion symmetry breaks results in drastic modification of the shape of the FMR absorption trace. The three involved symmetry breaks are: the layered character of the nanostructure with different materials for the two layers, the $L$-shape for the nanostructure cross-section and the single-side incidence of the driving microwave field.

From this point of view it is instructive to compare the Figs. 2(e1) with Fig. 4(b) from [44]. The latter figure demonstrates am experimental stripline FMR response for L-shape stripes with similar sizes, but fabricated from a single material. The relative amplitudes of the higher-order FMR peaks are much smaller in the figure from [44]. Furthermore, in that figure the peak amplitude drops monotonically with an increase in the peak number. Hence, the combined effect of the three symmetry breaks for the layered stripes is much stronger than the effect of two symmetry breaks - the L-shape in combination with the shielding effect - for the single-material stripes. (Note that one may expect a weaker impact of the shielding effect on the FMR amplitudes for the sample from [44] because of uniformity of electric conductance across the stripe cross-section.)

The same considerations apply to Fig. 2(e2). In this case the thickness of the Fe layer is twice smaller than for Fig. 2(f2). This results in a somewhat more uniform profile of the microwave magnetic field over the cross-section of the sample geometry. This is consistent with reduction in the shielding effect with reduction in the thickness of the highly conducting iron layer overlaying the less-conducting Permalloy one. However, despite of the reduction in the shielding, its impact on the peak amplitudes remains strong - a trial simulation assuming a spatially uniform microwave magnetic field reduces the relative amplitudes of the higher-order modes by at least 2 times with respect to ones shown in Fig. 2(e2).

Thus, our theory is able to explain main features seen in the experimental data. It is worth noticing, however, that the important feature of the experimental traces in Figs. 2(e1) and 2(f1) – the presence of two large-amplitude higher-order modes (correspond to Peaks 3 and 4 in the respective theoretical data), with Peak 3 just slightly smaller than Peak 4, – does not come out naturally from our simulations. We mean that this shape is not observed in every simulation. One has to play around with the value of the saturation magnetization for the Fe layer and the exchange constant for the Permalloy layer in order to make the higher-frequency peak - Peak 4 - taller than the lower-frequency peak - Peak 3. Noteworthy is that for $L_1$=10nm and $L_1$=20nm we need different values of the exchange constant for Permalloy to generate the correct peak structure.

On the contrary, since in the experiment the same peak structure is repeated for both Fe layer thicknesses one may conclude that it is natural for this layered system. We claim that we do not observe this property in the simulation, because the theory does not take into account some material parameters which affect the experimental behavior. This results in the "renormalization effect" which we discussed in the beginning of this section. If those missing parameters were taken into account in the model, the correct double peak structure would come out naturally and for the same value of the exchange constant for Permalloy in both cases of iron layer thickness.

One also has to admit that our theory is unable to identify the small-amplitude peak seen in Fig.2(f1) to the left of the fundamental peak – at $f$=2.6 GHz. Just from the very general considerations (see Sub-section 2-B) it follows that the assumption of strong exchange coupling between the layers excludes existence of resonance modes with eigen-frequencies below the one for the fundamental acoustic mode. Still, we believe that our theory is correct, because it also able to explain the resonance peak linewidth. The next sub-section deals with this aspect.

### D. Computational results: FMR peak linewidth

From comparison of Figs. 2(c1) and 2(d1-f1) one sees that the resonance line of the fundamental mode for the single-layer iron stripes is much broader than for the Permalloy stripes. Iron is known to possess noticeably larger magnetic losses than Permalloy, therefore the larger linewidth. There is no noticeable difference in the widths of the resonance peaks from Figs. 2(b1) and 2(c1). Thus, the tilted deposition of the iron layer does not deteriorate its growth. Interestingly, the linewidth shrinks a lot when the iron layer is deposited on top of the Permalloy one (Fig. 2(d1)), as we already mentioned in the preceding sub-section.

The peak widths for the simulated data were extracted in the same way, as for the experimental data – by fitting the computed traces with Eq.(1). Then the Gilbert damping parameters for the layers were varied and the simulations were repeated till we obtained the simulated peaks widths close to the experimentally measured ones.

The values of the Gilbert damping parameters for the layers which results in the best agreement of the simulated peak widths with the experiment are shown in the last column of Table 1. As follows from the table, one needs to assume that the Gilbert damping constants are the same for both layers and much smaller than for the reference single-layer stripes in order to obtain the narrow resonance lines for the peaks from Fig. 2(d2).

Noteworthy is the fact that fitting the resonance lines for the L-shape stripes requires larger values of the damping constant. Interestingly, similar to Fig. 2(d2) one needs to assume the same value of the damping constant for both layers in order to fit the lines from Fig. 2(e1). The increase in the iron layer thickness (Fig. 2(f2)) leads to a larger damping constant for the iron layer than for the Permalloy one.

This behaviour of the resonance linewidth may suggest that viscous damping given by the Gilbert term of LLG is not the only contributor to the damping. There may be at least one more strong contribution which is not localized in a particular layer, but is a joint property of both layers. Since the difference in the Gilbert constants increases with the increase in the thickness of the iron layer, one may suppose that scattering at the interface of the two layers is an important source of losses which meets the criterion of not being localized in a particular layer.

One more possible mechanism is scattering from the edge $x$=170nm which represents a magnetic non-uniformity affecting both layers. Hence, roughness of this edge will influence dynamics of both layers. The thicker the iron layer, the stronger the effect of the presence of the edge. The latter leads to larger contribution of the scattering from it to the total losses. This is consistent with an observation from Fig. 3 that the increase in the Fe layer thickness increases the resonance linewidth.

The difference in the damping constants for the two layers seen in the last row of Table 1 possibly tells us that for the 20nm-thick iron layer the contribution of the viscous damping in the that layer becomes an important part of the total losses.

Noteworthy is the behaviour of the peak widths as a function of the values of the Gilbert constant for the layers which our model demonstrates. Computations show that the width of Peak 1 is mainly determined by the loss constant for the Permalloy layer. The one for Peak 4 is more sensitive to $\alpha_G$ for the iron layer than for the Permalloy one. The width of Peak 2 is more or less equally affected by both constants.

This behaviour is consistent with the mode profiles from Fig. 5. For Peak 1 the magnetization dynamics is mostly concentrated in Permalloy, hence the energy loss due to viscous damping in Permalloy should be the dominating loss mechanism for this mode in the model. Mode 4 is characterised by large precession amplitude in the iron layer. Therefore losses in iron should strongly contribute to the total width of Peak 4. Thus, a gradual growth of the precession amplitude in the iron layer with an increase in the mode number explains the gradual increase in the impact of the loss constant for iron on the widths of resonance peaks observed as a function of the peak number in the model.

**IV Conclusion**

In this work we carried out systematic experimental and theoretical investigations of the ferromagnetic resonance (FMR) response of quasi-two-dimensional magnetic nano-objects – very long nanostripes made of ferromagnetic metals. We were interested in the impact of the symmetries of this geometry on the response. Three possible inversion symmetry breaks were of interest – (1) due to the shape of the stripe cross-section, (2) due to the double-layer structure of the stripes with exchange coupling between the layers, and (3) due to the single-side incidence of the microwave magnetic field on the plane of the nano-pattern. The latter is characteristic to the stripline FMR configuration.

It was found that the combined effect of the three symmetry breaks is much stronger than impacts of each of these symmetry breaks taken separately. The combined effect takes the form of very strong excitation of higher-order standing spin wave resonances in the layered stripes with an L-shape for the stripe cross-section. As a result, a non-monotonic behaviour of the resonance peak amplitude as a function of resonance number is observed. In the experimental traces this effect is seen as formation of a dense family of high-amplitude modes.

The developed theory revealed that this is due to enhancement of the FMR response of a series of resonances which are mainly localised in the wider Permalloy layer, when the character of precession of magnetization in the narrower iron layer changes from forced to resonance one. This happens for a particular frequency range close to the frequency for the fundamental FMR mode of a single layer iron stripe with the same cross-section sizes as for the iron layer of the L-shaped stripes.

The simulations also revealed a strong effect of microwave shielding by microwave eddy currents induced in these highly conducting stripes with doubly broken inversion symmetry of the stripe cross-section. A separate electromagnetic simulation demonstrated that the overlaying narrower but higher-conductivity iron layer strongly shields the microwave magnetic field in the underlying wider Permalloy layer. Importantly, the shielding effect strongly reduces the amplitude of the microwave magnetic field not only in the part of the Permalloy layer covered by iron, but also in its part which is free from the iron cap. The resulting strong non-uniformity of the total microwave magnetic field inside the stripes further enhances the FMR absorption amplitudes of the modes from the dense family of modes.

**Appendix. Calculation of the Oersted field of a microwave eddy current in a bi-layer metal stripe of rectangular cross-section**

In our calculation we assume that a spatially uniform microwave magnetic field is incident from $z=-\infty$ on the surface of a dense plane periodic array of bi-layer nano-stripes with $w_1=w_2=w$ (Fig. 1(a)) . We also assume that the density of the microwave eddy current induced in the stripes by the incident microwave magnetic field is uniform across the stripe system. This is a valid assumption since the stripe size is significantly smaller than the microwave skin depth for Permalloy (about 1 micron).

For simplicity we also assume that the eddy-current density is the same as for a continuous film of the same thickness and the same conductivity. Let us set the amplitude of the incident field to 1. Then from Eqs.(B8) and (B13) in Ref.[46] it follows that the eddy current densities $j_1$ and $j_2$ in Layers 1 ($0<z<L_1$) and 2 ($L_1<z<L_1+L_2$) respectively obey the following relations:

$$\begin{aligned} j_1 &= \frac{2z_0\sigma_1}{2+z_0(\sigma_1 L_1+\sigma_2 L_2)} \\ j_2 &= \frac{2z_0\sigma_2}{2+z_0(\sigma_1 L_1+\sigma_2 L_2)} \end{aligned}, \qquad (7)$$

where $z_0$ is the characteristic impedance of the surrounding space (e.g. 377 Ω for vacuum) and $\sigma_1$ and $\sigma_2$ are electric conductivities of Layers 1 and 2 respectively.

By using the Biot-Savarre-Laplace law we then arrive to the following formula for the $x$-component of the Oersted field in Layer 1:

$$\widehat{h}_x^{(1)}(x,z)=\frac{1}{2\pi}\sum_{i=-\infty}^{\infty}\left[g_x^{11}(x-i(w+d))\,j_1+g_x^{12}(x-i(w+d))\,j_2\right], \quad (8)$$

where

$$\begin{aligned} g_x^{11}(x,z) &= \frac{-1}{2}\int_0^w \ln\frac{z^2+(x-x')^2}{(z-L_1)^2+(x-x')^2}dx' \\ g_x^{12}(x,z) &= \frac{-1}{2}\int_0^w \ln\frac{(z-L_1)^2+(x-x')^2}{(z-L_1-L_2)^2+(x-x')^2}dx' \end{aligned}, \qquad (9)$$

and $d$ is the edge-to-edge distance between neighbouring stripes.

Similar expressions are obtained for the $y$-component of the Oersted field and for the fields in Layer 2. Note that the integrals in Eq.(9) exist in closed form. However, while using MathCAD software it is easier to calculate them numerically with tools for numerical integration built in into MathCAD than to derive those cumbersome analytical expressions and enter them into the MathCAD worksheet. Also note that the calculated field has collective character – for each of the stripes it is a sum of the field produced by the eddy current in the stripe and of the fields of all its neighbours at the position of the stripe (see the summation sign in Eq.(8).)

An example of calculation using Eq.(8) is shown in Fig. 6. One sees significantly larger amplitude of the microwave magnetic field in the iron layer with respect to the Py one. This difference in amplitudes is an additional factor resulting in high excitation amplitudes of the optical mode for the bi-layer stripes (see the dashed and dotted lines in Fig. 2(d2)).

**Acknowledgement**

Financial support by the Ministry of Education of Singapore, the Australian Research Council, the University of Western Australia (UWA) and the UWA's Faculty of Science is acknowledged. I.S.M. is supported by an UPRF fellowship from UWA and the ARC Centre of Excellence for Nanoscale BioPhotonics (CNBP).

Table 1.

| Sample | Number of layers and used materials | Layer width (nm) | Layer thickness (nm) | Gyromagnetic coefficient (MHz/Oe) | Saturation magnetization $4\pi M$ (G) | Exchange constant $A$ ($10^{-6}$ erg/cm) | Interlayer exchange constant $A_{12}$ (erg/cm$^2$) | Gilbert damping parameter |
|---|---|---|---|---|---|---|---|---|
| A | 1, Py | 340 | 10 | 2.9 | 7000 | 1 | N/A | 0.017 |
| B | 1, Fe | 340 | 10 | 3.05 | 12500 | 2 | N/A | 0.042 |
| C | 1, Fe | 170 | 10 | 3.05 | 13000 | 2 | N/A | 0.041 |
| D | 2, Fe/Py | 340/340 | 10/10 | 2.9/3.05 | 16500/10000 | 2/1 | 6 | 0.008/0.008 |
| E | 2, Fe/Py | 170/340 | 10/10 | 2.9/3.05 | 18500/7000 | 2/0.7 | 6 | 0.016/0.016 |
| F | 2, Fe/Py | 170/340 | 20/10 | 2.9/3.05 | 18500/7800 | 2/1.2 | 6 | 0.032/0.020 |

Table 1. Details of the samples used to produce experimental and theoretical data from Fig. 2. Columns 2 to 4: details of the sample geometries. Columns 5 to 9: values of magnetic parameters extracted from the best fits of the experimental traces with the theory. The parameters for the bi-layer samples are given in the format $t_1/t_2$, where $t_1$ and $t_2$ are the values for a parameter for the iron and Permalloy layers respectively.

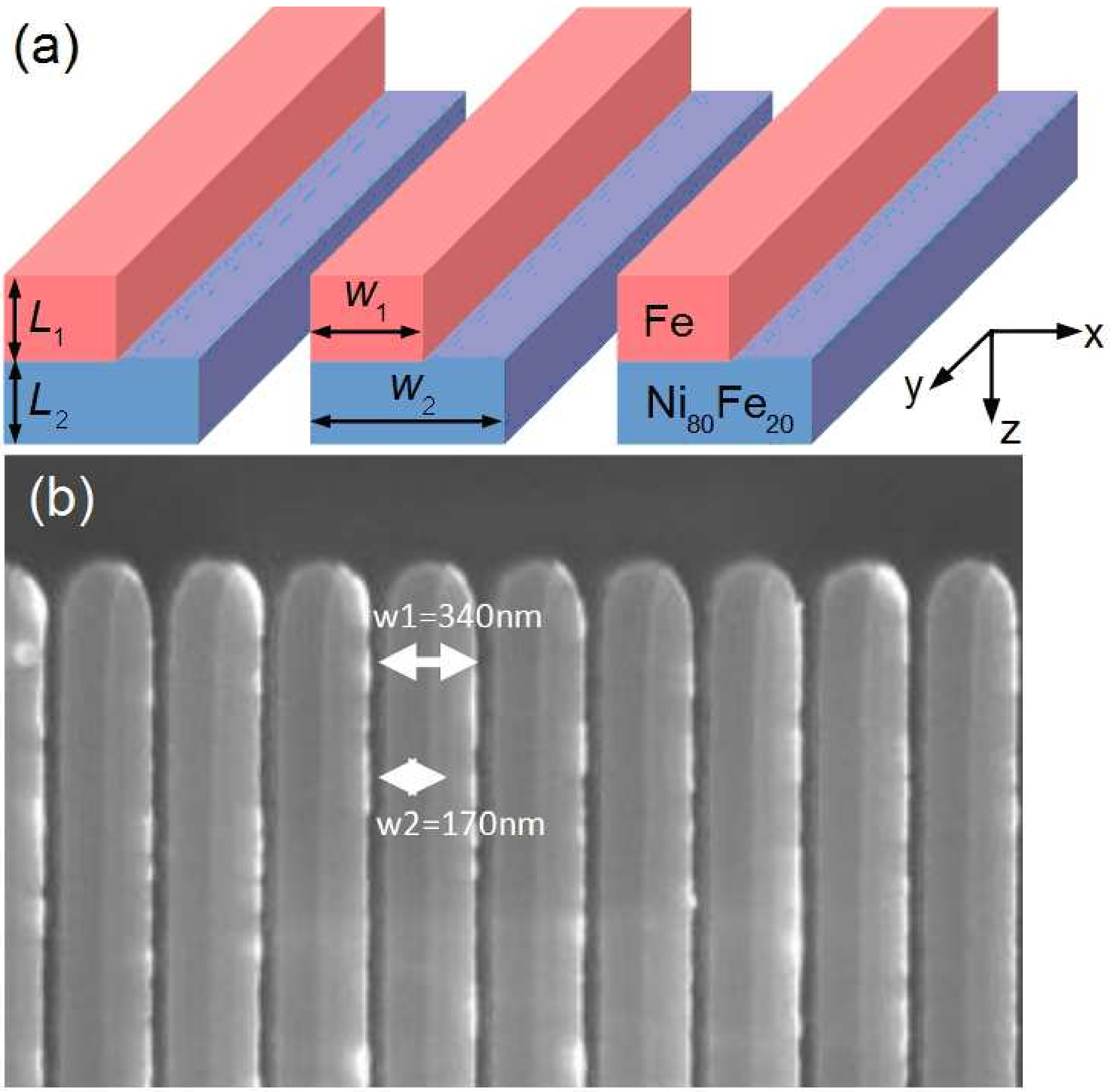

Fig. 1. (a) Sketch of the geometry of the L-shaped stripes. (b) SEM image (top view) of one of the fabricated nanostructures.

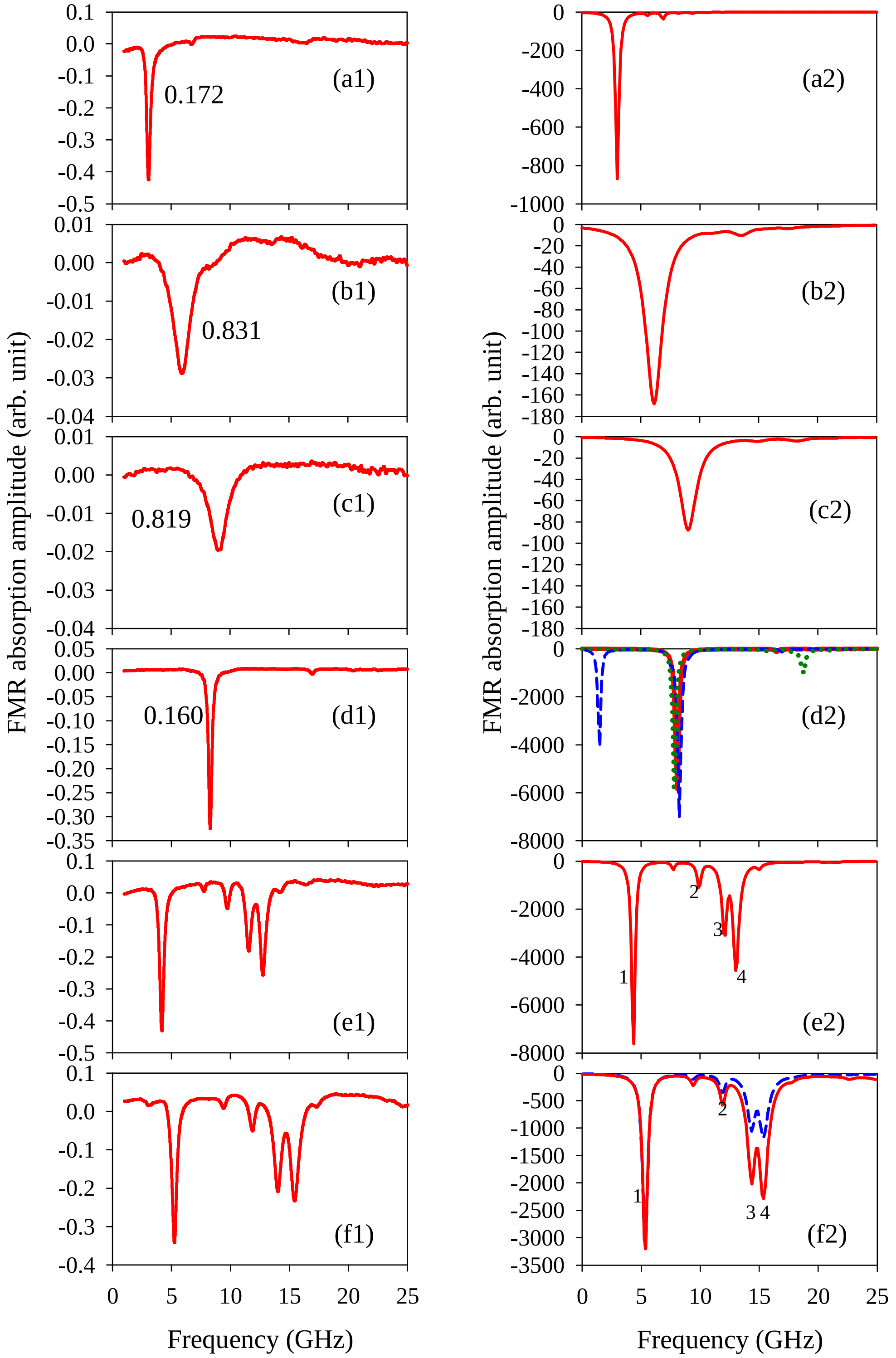

Fig. 2. Left-hand column: experimentally measured FMR traces. Right-hand column: simulated traces. Panels (a) to (f) correspond to samples A to F from Table 1. Solid lines in the right-hand panels: simulations for the input parameters shown in Table 1 and the microwave shielding effect included. Dashed and dotted lines in Panel (d2): interlayer exchange constant $A_{12}$=0 and 0.2 erg/cm$^2$ respectively. Dashed line in (f2): FMR is driven by spatially uniform microwave magnetic field (i.e. the shielding effect is not included.) Figures in (a1-d1): widths of respective resonance peaks (in GHz) extracted by fitting these traces with Eq.(1).

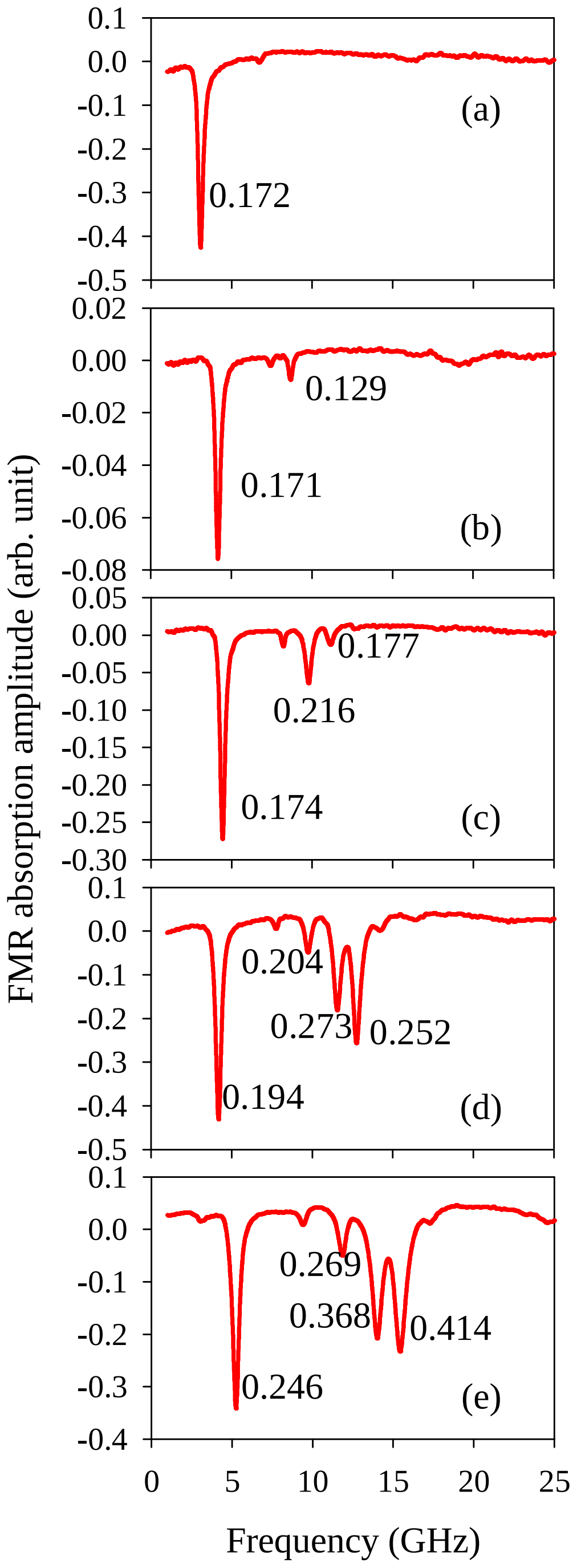

Fig. 3. Evolution of the FMR response of the L-shaped stripes with an increase in the thickness of the iron layer $L_1$. (a): no iron layer (the same as Fig. 2(a1)); (b) $L_1$=2nm; (c) $L_1$=5nm; (d) $L_1$=10nm; (e) $L_1$=20nm. (The two bottom panels are repeats of Fig 2(e1) and Fig. 2(f1)). Figures in the panels show the widths of the respective peaks in GHz. The peak widths were extracted by fitting these traces with Eq.(1).

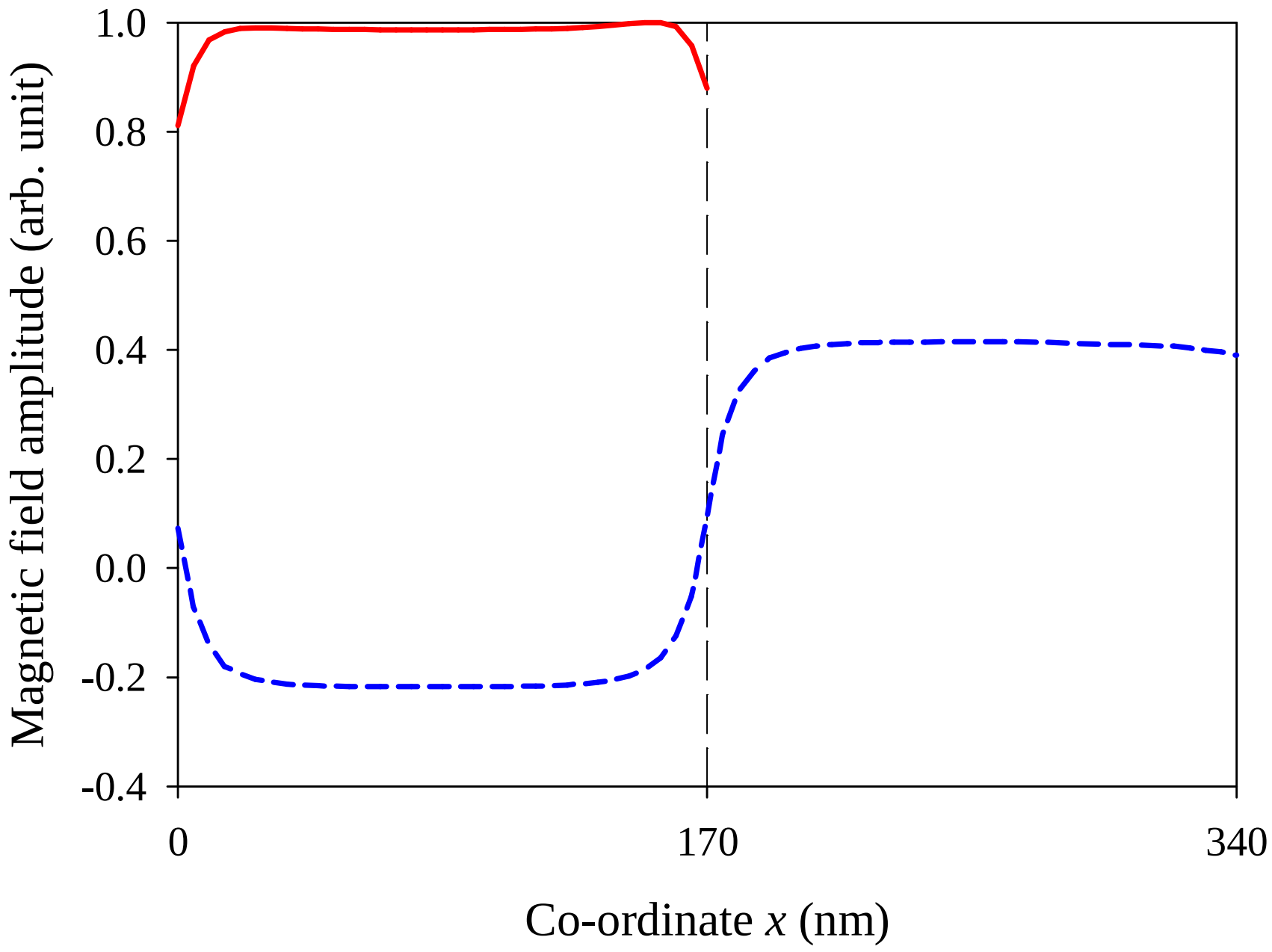

Fig. 4. Simulated in-plane ($x$) component of the total microwave magnetic field for the L-shaped bi-layer stripes with $L_1$=20nm and $L_2$=10nm. Solid line: the field inside the iron layer. Dashed line: the field inside the Permalloy layer. The fields are averaged over thicknesses of the layers. The total field is a sum of the microwave field of the stripline and of the field of the microwave eddy currents induced in the stripes by the field of the stripline. The vertical dashed line shows the position of the edge of the iron layer. The field amplitude is normalized to the maximum for the iron layer. Conductivity of the iron layer: $1.04\times10^7$ S/m; conductivity of the Permalloy layer: $4.5\times10^6$ S/m. The layers are treated as non-magnetic in this calculation.

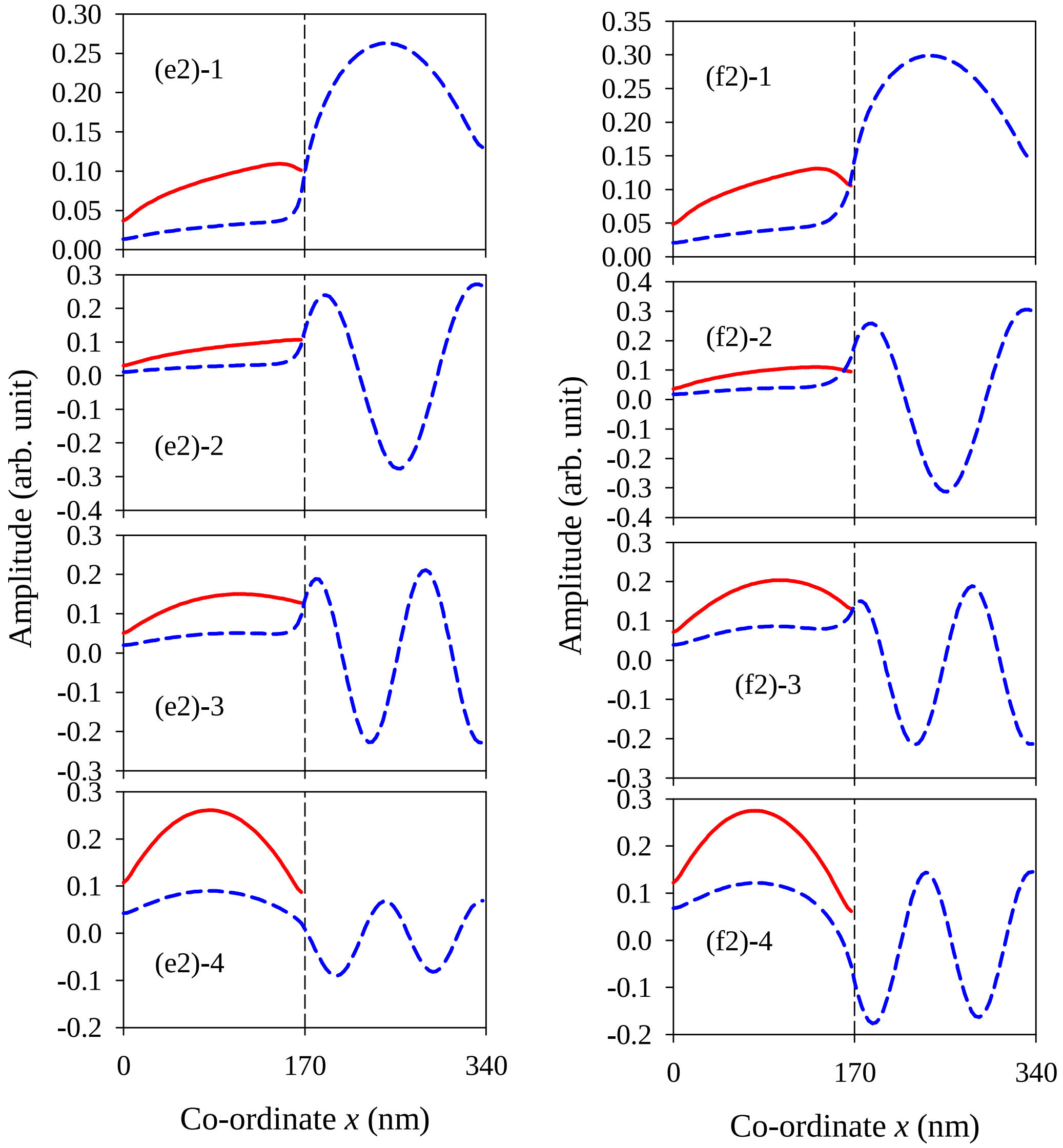

Fig. 5. Simulated profiles of the in-plane ($x$) component of dynamic magnetization for the modes responsible for Peaks 1 to 4 in Figs. 2(e2) and 2(f2). Solid lines: profiles for the iron layer. Dashed lines: ones for the Permalloy layer. The profiles are averaged over thicknesses of the layers. The parts of the panel numbers in the brackets - (e2) or (f2) - denote the respective panels of Fig. 2. The last digit of the panel numbers - 1, 2, 3 or 4 - denotes the respective peak in the respective panel of Fig. 2. The vertical dashed lines show the position of the edge of the iron layer.

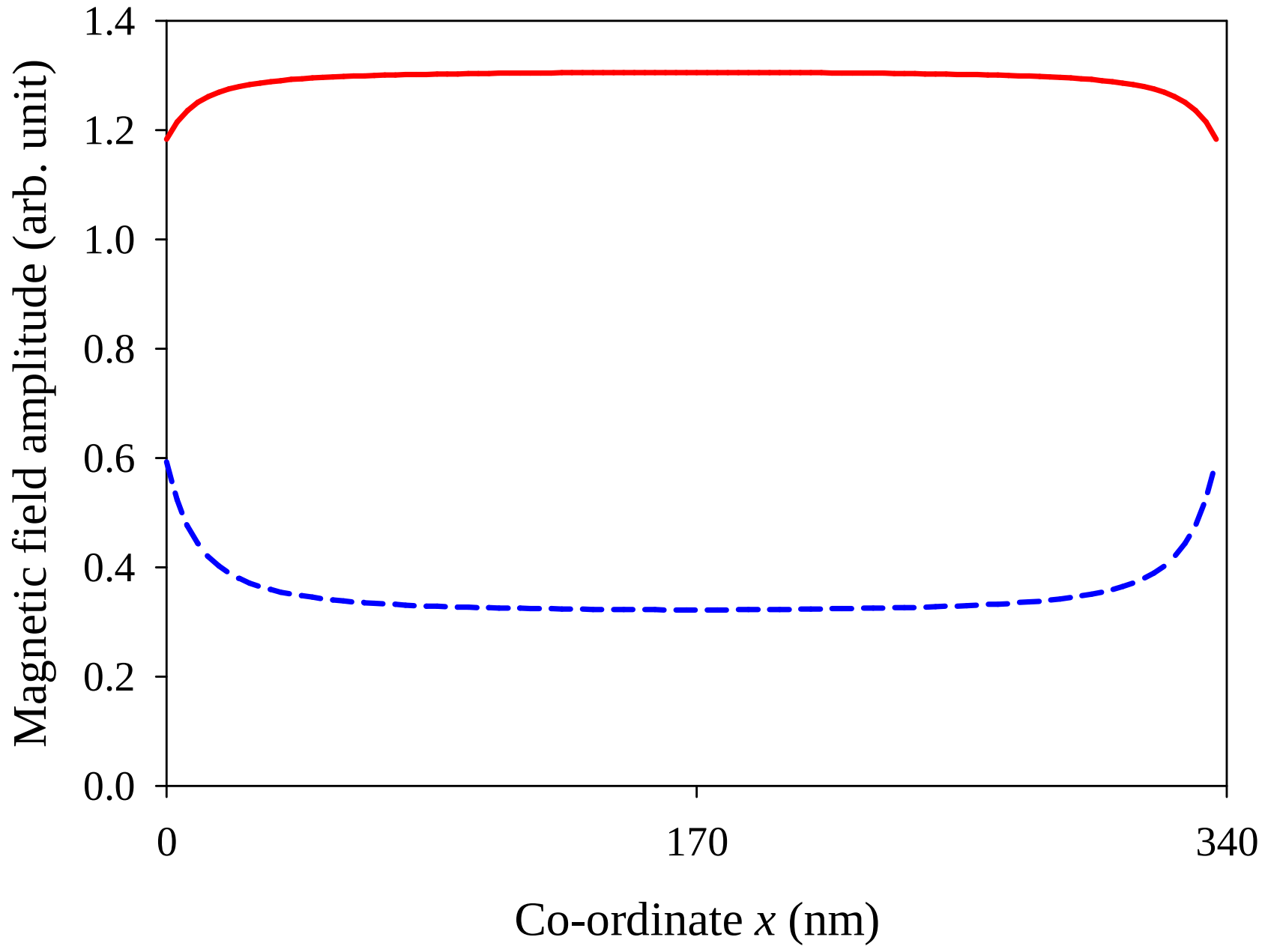

Fig. 6. In-plane (x) component of the total microwave magnetic field for the bi-layer stripes with rectangular cross-section shape ($L_1$=$L_2$=10nm; $w_1$=$w_2$=340nm). Solid line: the field inside the overlying iron layer. Dashed line: the field inside the underlying Permalloy layer. The fields are averaged over thicknesses of the layers. The amplitude of the incident microwave magnetic field of the stripline is 1. This calculation was carried with Eq.(8). Conductivities for iron and Permaloy layers: $1.04\times10^7$ S/m and $4.5\times10^6$ S/m respectively.